\documentclass[aps,prl,twocolumn,floatfix]{revtex4-1}
\usepackage{graphicx,epsfig,array,amsmath,hyperref}

\begin{document}

\title{Emergence of a Novel Pseudogap Metallic State in a Disordered 2D Mott Insulator}

\author
{Elias Lahoud$^{1}$, O. Nganba Meetei$^{2}$, K.B. Chaska$^{1}$, A Kanigel$^{1}$,  Nandini Trivedi$^{2\ast}$}

\affiliation{$^{1}$Physics Department, Technion-Israel Institute of Technology, Haifa 32000, Israel \\
$^{2}$Department of Physics, The Ohio State University, Columbus, Ohio 43210, USA}

\begin{abstract}
We explore the nature of the phases and an unexpected disorder-driven Mott insulator to metal transition in a single crystal of the layered dichalcogenide 1T-TaS$_{2}$ that is disordered without changing the carrier concentration by Cu intercalation. Angle resolved photoemission spectroscopy (ARPES) measurements reveal
that increasing disorder introduces delocalized states within the Mott gap that lead to a finite conductivity, challenging conventional wisdom. Our results not only provide the first experimental realization of a disorder-induced metallic state but in addition also reveal that the metal is a non-Fermi liquid with a pseudogap with suppressed density of states that persists at finite temperatures. Detailed theoretical analysis of the two-dimensional disordered Hubbard model shows that the novel metal is generated by the interplay of strong interaction and disorder.

\end{abstract}

\maketitle


The interplay of disorder and strong correlations and the resulting quantum phase transitions are long standing problems and continue to be major areas of research in condensed matter physics today \cite{abrahams_RMP_2001,kravchenko_2004,kravchenko_1994,punnoose_2005,vojta_1998,vlad1,vlad2,heidarian_2004,nandini_book}.  
Mott insulators at commensurate filling emphasize the role played by repulsive interactions in localizing electrons. These insulators have a gap to charge excitations and are often associated with magnetic order and low lying spin excitations\cite{fazekas_book,fujimori,antoine}. On the other hand, Anderson insulators describe non-interacting electrons in a random potential which show a mobility gap but have gapless single particle excitations and no magnetic properties \cite{abrahams_1979,lee_1985}.
The overarching question we ask in this article is the effect of disorder on the charge and spin excitations of a Mott insulator as illustrated in Fig. 1. A major challenge is to find strongly correlated quantum materials where disorder can be introduced in a controlled manner without changing the carrier concentration. The layered dichalcogenide 1T-TaS$_{2}$ with Cu intercalation is an excellent candidate for exploring precisely this interplay between strong correlations and disorder. 

\begin{figure}[htb!]
\includegraphics[trim=2cm 4cm 2cm 3cm,width=0.4\textwidth]{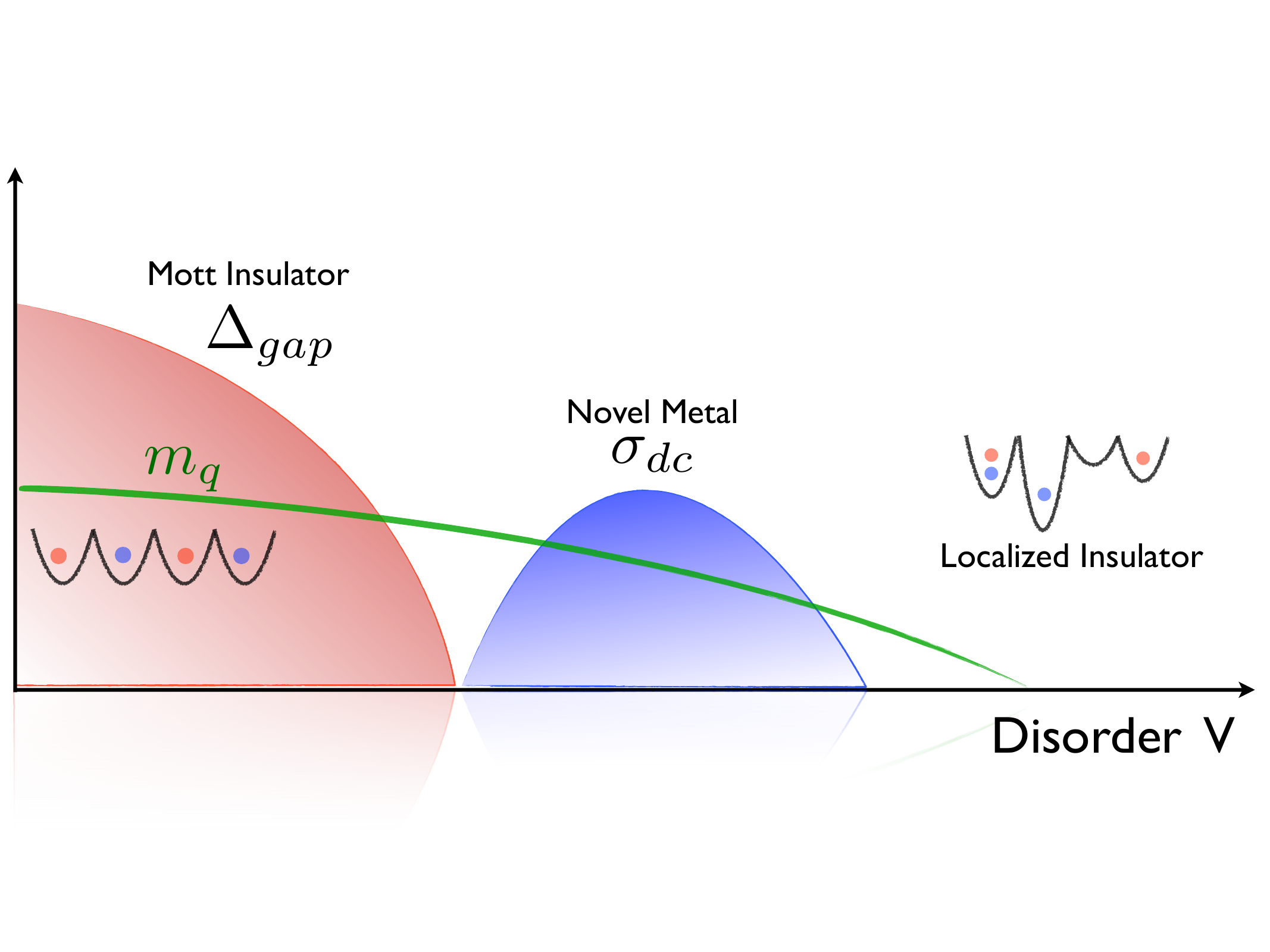}
\caption{Schematic figure showing the effect of disordering a Mott insulator (without adding carriers). As the disorder strength $V$ is increased relative to the bandwidth, the charge gap $\Delta_{gap}$ in the Mott insulator vanishes at the quantum phase transition to a metal. The novel metal has a finite dc conductivity $\sigma_{dc}$ and other unusual spectral properties elucidated by ARPES experiments reported here. At higher disorder we expect a second quantum phase transition from the metal to a gapless localized insulator. If the Mott insulator has magnetic order $m_q$, it need not be tied to the metal insulator transitions.}
\label{Fig1}
\end{figure}

The transition-metal dichalcogenide 1T-TaS$_2$ is a quasi 2D system with a very rich phase diagram. Upon decreasing the temperature it undergoes several charge density wave (CDW) phase transitions \cite{wilson_1975}. It even becomes superconducting when subjected to external pressure or chemical doping \cite{sipos_2008,grioni_2010}. 

In this joint experiment and theory collaboration, we focus on the Commensurate CDW (CCDW) phase below T=180K, shown to be a Mott insulator, and 
explore the effect of disorder without doping introduced by intercalating Cu.  Using detailed ARPES data along with transport measurements we show that a novel disorder induced metal is realized. It has surprising non-Fermi liquid behavior characterized by a temperature dependent pseudogap in the spectral function as shown in Fig. 3. Our theoretical analysis of the disordered Hubbard model reveal the origin of the extended states in the Mott gap and the pseudogap in the spectral function as summarized in Fig. 4.  
 
\begin{figure*}[ht!]
\includegraphics[trim=2cm 5cm 2cm 5cm,width=0.75\textwidth]{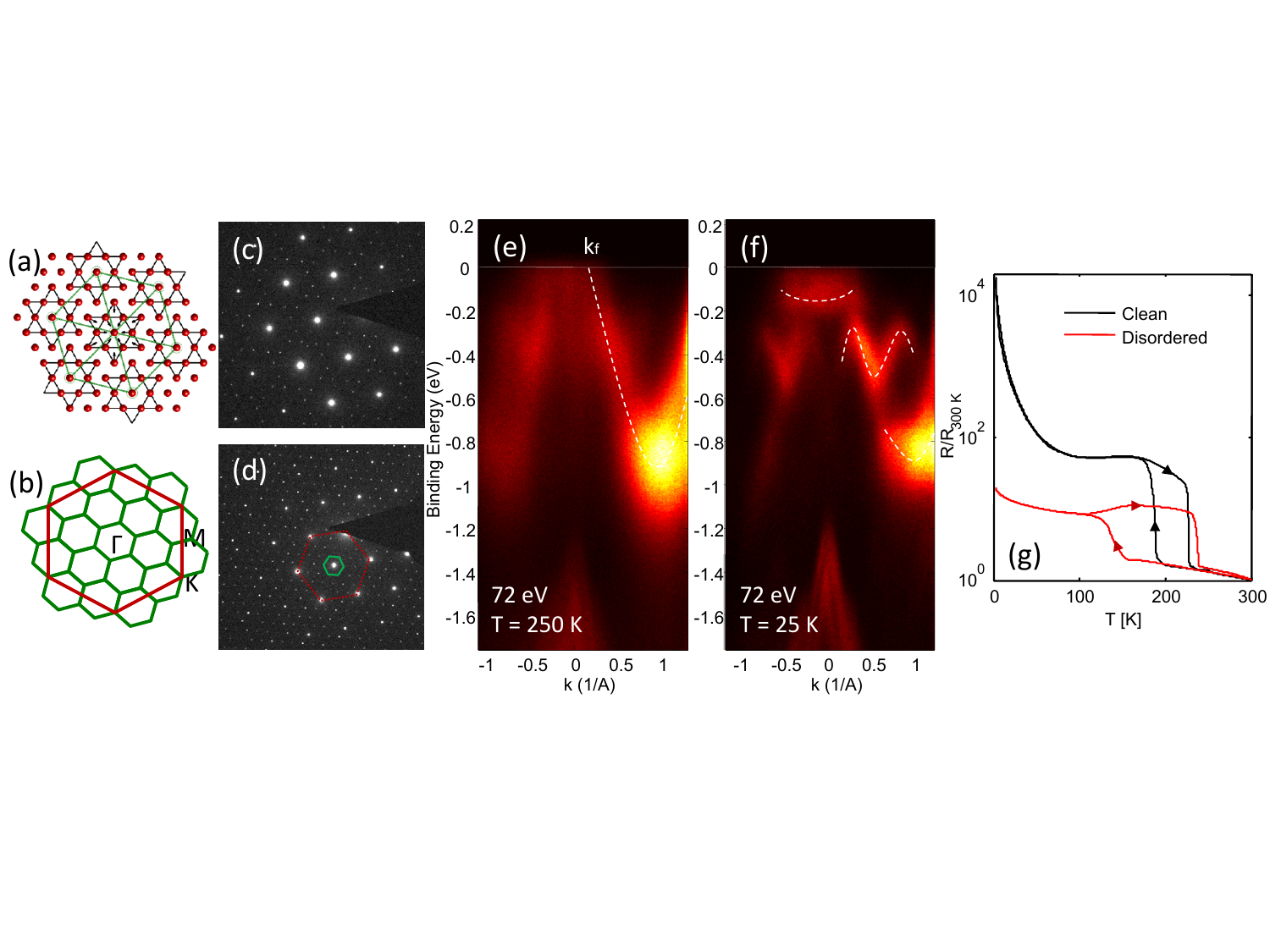}
\caption{Mott phase in 1T-TaS$_2$: (a)  visualization of the CCDW phase and star-of-David clusters on a triangular lattice. (b)  The 2D Brillouin zone of the original triangular Ta sheet (red), and the CCDW phase reconstructed reciprocal unit cells (green). (c-d) TEM diffraction images from a disordered sample in the nearly commensurate CDW phase (T=223K) and the commensurate CDW phase (T=123K) respectively. (e-f) ARPES data from a disordered sample showing the band dispersion along the $\Gamma$-M direction above and below the CCDW phase transition at 180K respectively. The dashed lines are guide to the eye showing the band dispersion. (g) Inplane resistance as a function of temperature for a clean and a disordered sample upon cooling from 300K down to 2K and heating back to 300K. 1T-TaS$_2$ is quasi-2D with large resistance anisotropy ($\rho_{c}/\rho_{a}\approx 500$)\cite{rho1,rho2}.
For a discussion of the unusual temperature dependence of the resistivity, see supplement~\cite{supp}.
}
\label{Fig2}
\end{figure*}

The pseudogap, a state showing suppression in the density of states, has emerged as a new state of matter in very different systems though suggesting very different origins. By now it has been shown to arise due to pairing correlations persisting in the normal state \cite{mohit_1992,campuzano,gaebler}, due to competing spin and charge orders \cite{yazdani}, as well as due to disorder-generated hard gap and pseudogap in superconducting thin films \cite{ghosal_prl,ghosal_prb,karim,sacepe1,sacepe2}. Our discovery here is significant because it points to yet another factor, Coulomb correlations and Mott physics, driving the dominant mechanism for a pseudogapped metallic state in Cu-intercalated 1T-TaS$_2$.


\emph{1T-TaS$_2$ (Single band Mott insulator):} We start by establishing that 1T-TaS$_2$ is a single orbital Mott insulator. Despite the seemingly complicated electronic properties, 1T-TaS$_2$ has a simple crystal structure composed of weakly coupled layers, each layer containing a single sheet of tantalum atoms, sandwiched in between two sheets of sulfur atoms. The Ta atoms within each layer form a 2D hexagonal lattice. The basic CDW instability is formed within the tantalum layers by the arrangement of 13 Ta atoms into a ``star-of-David" shaped cluster (Fig. 2(a)). As temperature is lowered, the size of the CDW domains becomes larger until all the domains interlock into a single coherent CDW modulation  extending throughout the layer in the CCDW phase below T$_{CCDW}$=180K. The new ``star-of-David" unit cells form a hexagonal lattice and the corresponding reduced Brillouin zone is depicted by the small green hexagon in Fig. 2(b).

The transition into the CCDW phase is accompanied by a metal-insulator (MI) transition into a Mott phase \cite{wilson_1975,fazekas_1980}, clearly seen from resistivity measurements shown in Fig. 2(g). 
In agreement with previous works~\cite{whangbo_1992}, the structural CCDW formation also leads to an electronic reconstruction
as seen from the ARPES spectra. Before reconstruction a parabolic band (dashed line in Fig. 2(e)) with a bandwidth $\approx$ 1eV, originates from the Ta 5d orbital\cite{smith_2006}, 
and disperses upwards from M towards the $\Gamma$ point, crossing the Fermi energy at ~25 \% of the $\Gamma-M$ distance. After CCDW reconstruction, the upper most sub-band, laying closest to the Fermi energy, is well separated in energy from the rest with a bandwidth of about $W\approx$ 45meV. The Coulomb on-site energy $U$  is evaluated to be on the order of 0.1eV \cite{sipos_2008}. 
Thus the low energy physics of 1T-TaS$_2$ is captured by a 2D single-band Hubbard model on a triangular lattice~\cite{fazekas_1980} with the value of $U/W \approx 2.2$ well over the critical value for the Mott transition($U_C/W \approx 1.3 $)\cite{scalettar_2006}. 

\begin{figure*}[ht!]
\includegraphics[trim=2cm 5cm 2cm 6cm,width=0.8\textwidth]{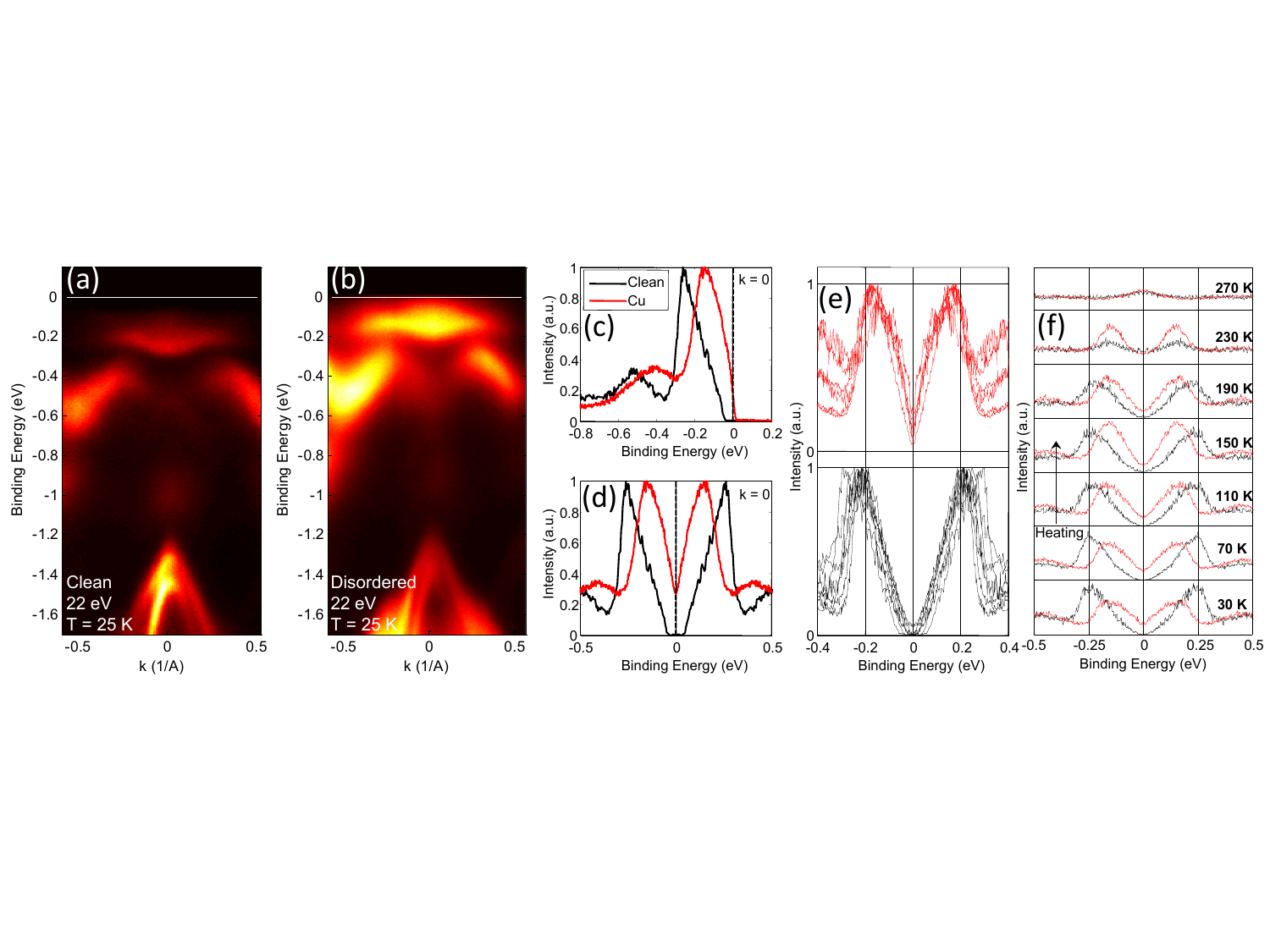}
\caption{ARPES spectra of a disordered Mott-insulator 1T-TaS$_2$: Spectra taken with 22eV photons  along the $\Gamma-K$ direction at T$=25K <{\rm T}_{CCDW}$. Panel (a) for a clean system reveals a Mott gap as indicated by the lack of intensity within about 80 meV of the Fermi energy. Panel (b)for a disordered sample shows the presence of significant spectral weight close to the Fermi energy,  inside the energy gap. In panels (c-f), the scans in black are for a clean system while those in red are for a Cu intercalated disordered system. (c)  Comparison of the EDC at the $\Gamma$ point taken from the scans in (a) and (b). The peak in the disordered 1T-TaS$_{2}$ is broadened and shifted towards lower binding energy and there is substantial intensity at zero energy. (d) The same EDCs shown in (c) after symmetrization clearly reveals the closing of the gap in the disordered sample, and the formation of a pseudogap with about 20 \% of the peak-intensity. (e) Reproducibility of the results: EDCs from 13 different samples (7 clean and 6 disordered) taken in different ARPES systems with different photon energies and polarizations, all reveal the same qualitative behavior. (f)  Evolution of the pseudogap at the $\Gamma$ point as function of increasing temperature. While at low temperatures the spectral weight at zero energy is much larger in the disordered samples, above the CCDW transition temperature the difference in the spectral functions of the clean and disordered samples disappears.}
\label{Fig3}
\end{figure*}

\emph{Cu intercalation:} 
The presence of intercalated Cu can locally deform the crystal and the long-range strain fields\cite{frenkel_2006} can lead to modifications of the electronic structure of the CCDW phase. We have modeled the effects of randomly distributed Cu on the CCDW phase by a disordered Hubbard model. The CCDW phase remains robust against this disorder~\cite{supp} as seen from 
(a) the TEM images showing similar diffraction patterns in the clean and disordered systems, (b) the large hysteresis in the temperature-dependent resistivity with a sharp transition, and (c) the formation of sub-bands due to Fermi surface reconstruction that persists upon intercalation. 

\emph{Spectral properties of a disordered Mott insulator:} ARPES data on disordered 1T-TaS$_2$ (see Fig. 3(a-b)) shows a significant increase of spectral weight in the $[-0.1 ~eV,~ 0 ~eV]$ energy range at the $\Gamma$ point. 
The energy distribution curves (EDCs) in Fig. 3(c) shows the transfer of spectral weight towards lower binding energy in the disordered sample, with a line shape that is slightly skewed towards E=0. 
The symmetrized EDC (Fig. 3(d)) confirms the presence of a ``soft" gap, where the spectral weight at zero energy is about 20 \% of the intensity at the peak and the broad peaks are shifted inwards, leaving a smaller Mott gap. The filling up of Mott gap is accompanied by an insulator to metal transition indicated by the drop in resistivity by several orders of magnitude (Fig. 2(g)). The emergent metallic phase exhibits novel non Fermi liquid characteristics like a distinct pseudogap around $\omega=0$ and an unusual low temperature resistivity that increases slightly with decreasing temperature (discussed in more detail later). The results are highly reproducible as seen in Fig.3(e). 


The pseudogap at low temperatures is found to persist up to higher temperatures. The temperature dependence of the symmetrized EDC at $\Gamma$ is shown in Fig. 3(f) for a clean and disordered sample. The difference in line shape and in the shape of the gap is most prominent at low temperatures, and the EDCs become increasingly similar as temperature is raised. 

The are several points about the data that are significant: (1) Unlike in semiconductors with rigid bands, we find no evidence of disorder generating significant amount of localized impurity states in the correlated Mott insulator. Such impurity states, if present, would have shown up above T$_{CCDW}$ as additional spectral weight near the chemical potential in the disordered sample compared to the clean case \cite{supp}, which is however not observed.(2) It is also clear that phase separation and domains of CCDW separated by metallic regions cannot explain the spectrum in the pseudogapped metal. The spectrum due to phase separation would be a weighted average of that from a Mott insulator and a normal metal, and as a result the Hubbard bands for the disordered system would remain fixed at the location of the clean system. This is inconsistent with our observations that the Hubbard bands in the disordered samples move to lower energies (Fig. 3(d)).

\emph{Disordered Hubbard model:} To gain better insight into our experimental results, we investigate the effect of disorder on a Mott insulator using a half-filled single band Hubbard model on a triangular lattice. It describes the CCDW phase of 1T-TaS$_2$. 
\begin{eqnarray}
H &=&-t\sum_{\langle ij\rangle ,\sigma}\left( c^{\dagger}_{i\sigma} c_{j\sigma} + h.c.\right) -t^\prime\sum_{\langle\langle ij \rangle\rangle}\left(c^{\dagger}_{i\sigma} c_{j\sigma} + h.c.\right) \nonumber \\
  && + \sum_{i,\sigma} \left(\epsilon_i-\mu\right)c^{\dagger}_{i\sigma}c_{i\sigma}+U\sum_i n_{i\uparrow}n_{i\downarrow} \label{eq:ham}
\end{eqnarray}
Here $t$ is the nearest-neighbor hopping between unit cells, $t^\prime$ the next nearest hopping and $U$, the on-site interaction strength. $c^{\dagger}_{i\sigma}(c_{i\sigma})$ creates(annihilates) an electron of spin $\sigma$ at site $i$, and $n_{i\sigma}=c^{\dagger}_{i\sigma}c_{i\sigma}$ is the number operator for spin $\sigma$ at site $i$. The onsite disorder potential  $\epsilon_i$ is chosen from a flat distribution in the range $[-V/2,V/2]$, and captures the effect of lattice distortions due to Cu intercalation. In our analysis, we tune the chemical potential, $\mu$, for each disorder realization to fix the density at half filling. We solve Eq.~\ref{eq:ham} within inhomogeneous mean-field theory. The interaction term is approximated as $n_{i\uparrow} n_{i\downarrow} \approx   n_{i\uparrow}\langle n_{i\downarrow} \rangle + \langle n_{i\uparrow} \rangle n_{i\downarrow} - \langle n_{i\uparrow} \rangle \langle n_{i\downarrow} \rangle -\langle S^{+}_i \rangle S^{-}_i - \langle S^{-}_i \rangle S^{+}_i + \langle S^{+}_i\rangle \langle S^{-}_i\rangle$ where the operator $\mathbf{S}=\sum_{\sigma,\sigma^\prime}c^\dagger_{\sigma}\mathbf{\tau}_{\sigma,\sigma^\prime} c_{\sigma^\prime}$ and $\mathbf{\tau}$ are the Pauli spin matrices. Due to the presence of disorder, the density and spin fields  are site-dependent and the $3N$ parameters (\{$\langle n_{i\sigma}\rangle, \langle S^{+}_i \rangle=\langle S^{-}_i\rangle^* \},\; i=1,\dots,N \; \textrm{and} \; \sigma=\uparrow or \downarrow$), where $N$ is the total number of sites,
are obtained by a self-consistent solution of the coupled inhomogeneous Hartree-Fock equations. 

At half-filling in the clean limit with $t^\prime =0$, mean field solutions gives an insulator for $U>U_c\simeq 5.27t$ \cite{krishnamurthy_1990} along with magnetic ordering of the spins on the triangular lattice in a commensurate $120^o$ pattern. For the rest of the calculation, we choose $U=6t$ in order to remain in the insulating phase. Inclusion of $t^\prime=0.3t$ does not affect the magnetic ground state, but is important for producing a realistic band structure for comparison with experiments. We also find that small changes in filling ($\delta n < 2\%$) does not change the mean-field solution. 

\begin{figure}[htb!]
\centering
\includegraphics[trim=2cm 7.5cm 2cm 6.5cm, width=8.5cm]{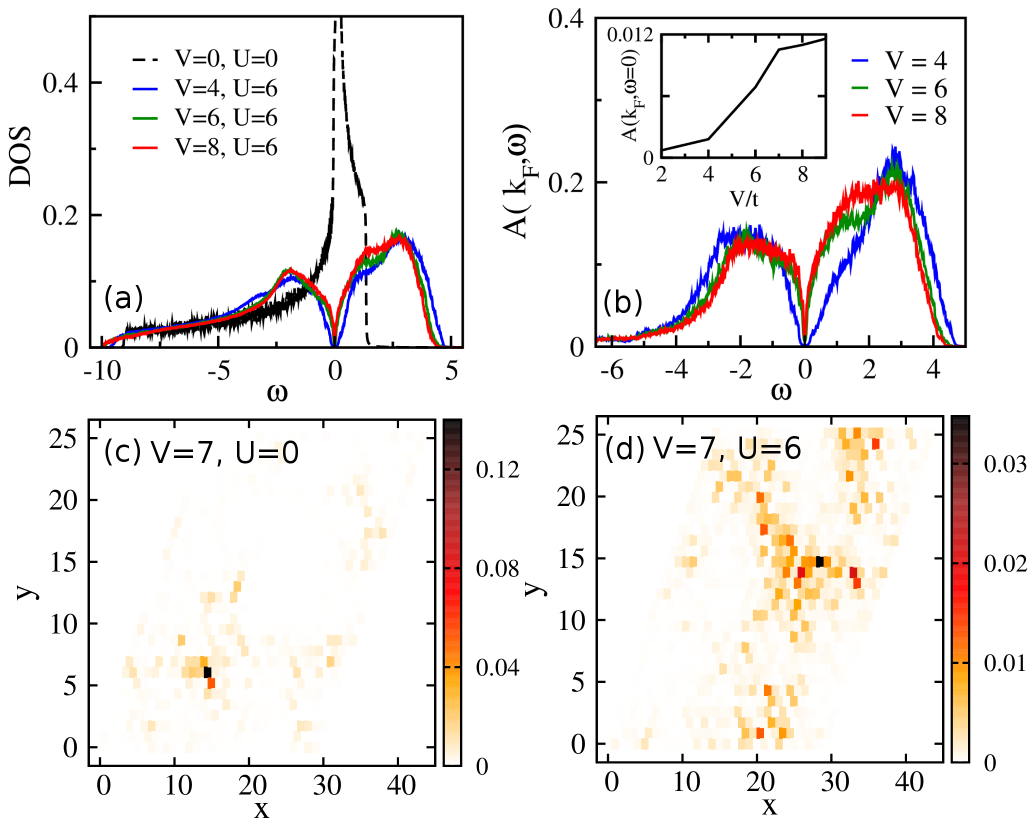}
\caption{(a) Theoretical density of states as a function of disorder $V$ (in units of hopping $t$) for interaction $U=6t$. (b) EDC at $k=k_F$ for $U=6t$ showing a well formed gap 
for weak disorder ($V<U$) which closes for $V\geq U$ with a pseudogap persisting.  Inset shows the monotonic increase of spectral weight at $\omega=0$ with disorder. (c) and (d) shows the spatial map of the probability density $| \Psi (i) | ^2$ of the eigenstate at $\omega=0$ for $V=7t$ and $U=0$ (panel (c)) and $U=6t$ (panel (d)). Remarkably repulsive interactions compete against disorder and delocalize the wavefunction over the entire lattice. 
The results in (a) and (b) are on 18x18 lattices averaged over 100 disorder realizations; (c) and (d) on a 32x32 lattice for a single realization.}
\label{Fig4}
\end{figure}

\emph{EDC of spectral function:}  For a given disorder realization, we numerically diagonalize the mean-field Hamiltonian exactly to obtain the single particle eigenstates $\{|\Psi_\alpha\rangle\}$ and eigenvalues $\{\epsilon_\alpha\}$ which are used to calculate the single particle Green's function 
\begin{equation}
G(\mathbf{r}_1,\mathbf{r}_2,\omega)=\sum_\alpha \frac{\langle \mathbf{r}_1|\psi_\alpha \rangle \langle \psi_\alpha | \mathbf{r}_2 \rangle}{\omega+i\eta-\epsilon_\alpha}
\end{equation} 
For a given disorder strength, $V$, we average over several disorder realizations, and then Fourier transform to get the spectral function $A(\mathbf{k},\omega)=-(1/\pi){\rm Im}\;G(\mathbf{k},\omega)$.

While previous studies have found a suppression in the total density of states around zero energy   ~\cite{chiesa_2008,imada_2009,wortis_2010} as also shown in Fig. 4(a), here our focus is on the disorder dependence of the spectral function as highlighted in Fig. 4(b). The EDC at the Fermi momentum for different amounts of disorder agree qualitatively with our experimental results shown in Fig. 3(c-d) on several aspects: 1) With increasing disorder the Mott gap closes and finite spectral weight develops at $\omega=0$ (see inset of Fig. 4(b)). 2) The shape of the gap changes from a hard `U' shaped gap in the clean limit to a `V' shaped soft gap in the disordered case. 3) The position of the broad peaks corresponding to the upper and lower Hubbard bands move closer to the chemical potential with increasing disorder in agreement with experimental data. In general, the spectral function is asymmetric around $\omega=0$ \cite{mohit_2005}. Since ARPES is only able to access the filled states, the experimental data has been symmetrized around $\omega=0$.

\emph{Transport properties in the pseudogapped metallic phase:} 
Within the relaxation time approximation, the conductivity is given by~\cite{supp}
\begin{equation}
\sigma_{xx}  =  2e^2 v_{F}^2 \tau_{F} \int d\epsilon g(\epsilon) \left(-\frac{\partial f_\epsilon}{\partial \epsilon}\right)
\end{equation}
where $g(\epsilon)$ is the density of states (DOS). A direct consequence of the strong energy-dependent $g(\epsilon)$ in the disordered Hubbard model is that the low temperature behavior of resistivity $\rho(T)$ increases with decreasing temperature and eventually saturates as $T\rightarrow 0$. This then defines the characteristic transport signature in the pseudogapped metallic phase (see Fig.~3 in the supplement \cite{supp}), distinct from a Fermi liquid that has a positive temperature coefficient of resistivity and from an insulator that shows no saturation at low temperatures. Such unusual low temperature behavior has also been reported in the context of high mobility suspended graphene samples with the chemical potential tuned close to
the Dirac point \cite{kim_2008}.

In conclusion, we have shown that while for non-interacting electrons disorder always localizes the wave functions, in strongly interacting systems, under certain conditions, disorder can {\it delocalize} them, quite unlike midgap states in a conventional semiconductor. 

This remarkable phenomena has been demonstrated here through ARPES measurements on Cu intercalated 1T-TaS$_2$.  We have established the first example of a non-Fermi liquid metal with a pseudogap generated upon disordering a 2D Mott insulator. In our calculation, the additional low frequency weight upon disordering the system arises primarily from k-states close to the underlying Fermi surface of the tight-binding model. In the experimental data in Fig. 3(b), on the other hand,  the additional weight appears from states close to the $\Gamma$ point. The discrepancy could be due to experimental resolution or possibly due to magnetic ordering and the resultant zone folding that can also push the weight close to the $\Gamma$ point. While many exotic phases have been found upon doping in proximity to Mott insulators and spin liquids such as the d-wave superconductivity in the cuprates and organics, in all these systems 
doping simultaneously introduces both carriers and disorder. We have explored a  regime where disorder is separately tuned without changing the carrier concentration. Our investigations open up the possibility of several future explorations of the nature of magnetism, its evolution with disorder and its relation to the metal-insulator transitions.

\noindent{\em{Acknowledgments:}} ONM and NT acknowledge support from DOE Grant No. DE-FG02-07ER46423 and the Ohio Supercomputing center for computational resources. The Synchrotron Radiation Center is supported by NSF DMR-0084402.

\bibliography{references}
\bibliographystyle{apsrev}
\end{document}


\title{Supplementary material for ``Emergence of a Novel Pseudogap Metallic State in a Disordered 2D Mott Insulator"}

\author
{Elias Lahoud$^{1}$, O. Nganba Meetei$^{2}$, K.B. Chaska$^{1}$, A Kanigel$^{1}$,  Nandini Trivedi$^{2\ast}$}

\affiliation{$^{1}$Physics Department, Technion-Israel Institute of Technology, Haifa 32000, Israel \\
$^{2}$Department of Physics, The Ohio State University, Columbus, Ohio 43210, USA}

\maketitle
In this supplement we give additional technical details about the experiment and present a theoretical understanding of the unusual resistivity of the pseudogapped metal.  

\section{Experimental details} 

\noindent \emph{\underline{Sample Preparation:}}
Single crystals of 1T-TaS$_2$ were prepared using the vapor transport method as described in Ref. \cite{prep}.  For the addition of disorder we intercalated Cu into the crystals. Not much is known about Cu intercalation in the 1T structure, but it was successfully done in 2H-TaS$_2$ \cite{2H-CuTaS2} and in 2H-TiSe \cite{ong_2006}.

For the intercalation, we added 10 \% Cu into the stoichiometric mix during the crystal growth stage. However, the copper content of the single crystals that emerge from the growth is well below 10 \%. A small increase in the C-axis lattice constant suggests that the Cu indeed resides in the Van-der-Walls gap \cite{2H-CuTaS2}. 

Electron back-scattering images  taken using a scanning electron microscope (SEM) reveal very uniform crystals. We used wave dispersive electron spectroscopy (WDS) to measure the Cu content. Again, we find a very uniform distribution of Cu, the average amount of Cu is found to be $0.75 \%  \pm0.05 \%$. 

 These samples, which we shall refer to as disordered 1T-TaS$_2$, display a lower transition temperature from a non-commensurate CDW to the commensurate CDW phase of around $150K$ upon cooling (Fig. 2(g)). The negative slope of the resistance $R(T)$ in Fig. 2(g) is also considerably reduced compared to the clean sample, suggesting enhanced conductivity. 
 
 The presence of intercalated Cu can produce several effects such as local deformation of the crystal and the generation of long-range strain fields \cite{frenkel_2006}. These structural variations can lead to modifications of the electronic structure of the CCDW phase. As discussed in the main text,  we have modeled the effects of randomly distributed Cu on the CCDW phase by a disordered Hubbard model. 

It has also been suggested that the addition of Cu could lead to domain formation. The transmission electron microscopy (TEM) images Fig. 2(c-d) scan a spot size of about a micron and show sharp diffraction spots from the locking of the ``star-of-David" clusters for the disordered Cu-intercalated 1T-TaS$_2$ above and below T$_{CCDW}$ that are identical to the clean sample~\cite{supp}. This suggests that if intercalation produces domains, they must be at least as large as the spot size. Furthermore, the reconstruction of the ARPES spectra 
(see Fig. 2(e-f)) from a parabolic band at high temperatures to a three sub-band structure due to the CCDW formation below 25K is seen in both clean (Fig. 3(a)) and disordered systems (Fig. 2(f))\cite{supp}, suggesting that the CCDW phase is robust.

We get further corroboration of the robustness of the CCDW phase from the temperature-dependent resistivity (Fig. 2(g)). As expected, it shows a large hysteresis that is characteristic of a first order phase transition into the CCDW phase. Of significance is the fact that the transition remains sharp and the hysteresis 
survives even in the disordered system. The hysteretic region is enlarged in the disordered case and could possibly arise from inhomogeneous regions having a variation in the transition temperature.

\noindent \emph{\underline{ARPES:}}
We have performed extensive angle-resolved photoemission spectroscopy (ARPES) measurements, on the PGM-beam line at the Synchrotron Radiation Center with $72~eV$ and $22~eV$ photon energy as well as at the ARPES machine at the Technion, using the $HeI~21.22~eV$ spectral line from discharge lamp as the light source. In both cases data was collected using a scienta R4000 electron analyzer. Samples were cleaved in-situ at a pressure better than $5\times10^{-11}$ torr. Energy and angular resolution were better than $10~meV$ and $0.3^{\circ}$ respectively.

\begin{figure}[htb!]
\centering
\includegraphics[width=8.5cm]{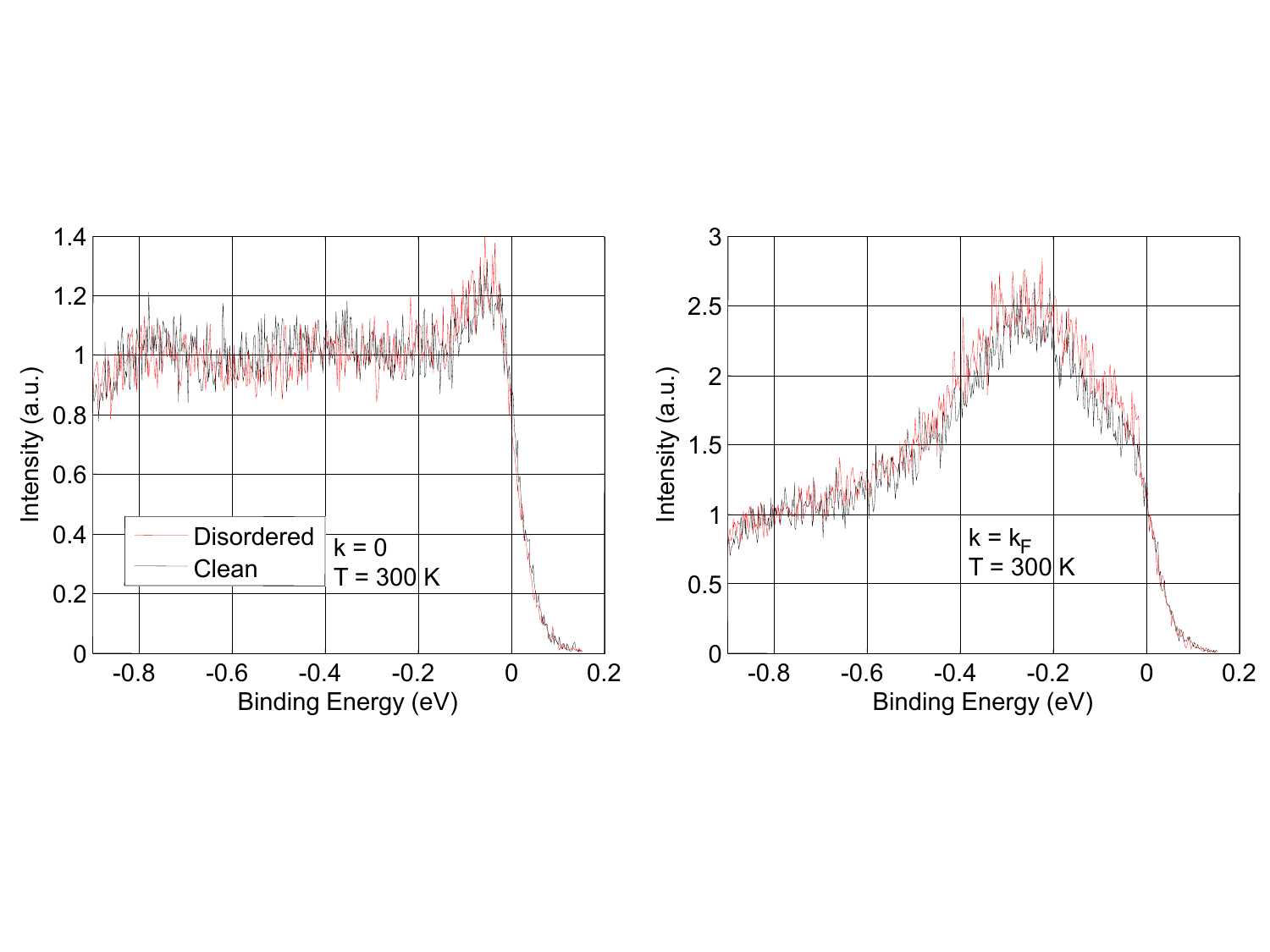}
\vspace{-1.5cm}
\caption{EDCs of a clean and disordered sample taken with $22~eV$ photon energy at room temperature at  $\Gamma$ in (a) and at $k_F$ in (b).}
\label{fig:EDCs_rt}
\end{figure} 

\bigskip

\noindent \emph{\underline{Disorder effects in the NCCDW:}}
We verify that the new states reported in the paper are linked to the underlying Mott state by searching for any sign of increased spectral weight near the Fermi level in the normal metal phase that emerges from the non-commensurate charge density wave (NCCDW) phase. In Fig.~\ref{fig:EDCs_rt} we compare the EDCs at the $\Gamma$ point and at k$_F$ of clean and a disordered sample. The samples were cleaved and measured at $300~K$ and the data normalized by the intensity at $-0.8~eV$. The EDC's are identical and this remains the case throughout the Brillouin zone. We find no trace of the new states when the disordered sample is not in the Mott state.
In addition, we measured very carefully the position of the top of the Sulfur band with respect to the Fermi-level. The top of Sulfur band at about 1.2eV below the Fermi-level is relatively sharp and can be used to measure the chemical-potential. We found no measurable effect of Cu intercalation on the chemical-potential.  
 We conclude that if indeed the Cu intercalation induces new states, these states cannot account for the additional spectral intensity within the Mott gap in the commensurate CDW (CCDW) phase. 

 \bigskip
 
\begin{figure}[t]
\centering
\vspace{-0.5cm}
\includegraphics[width=9cm]{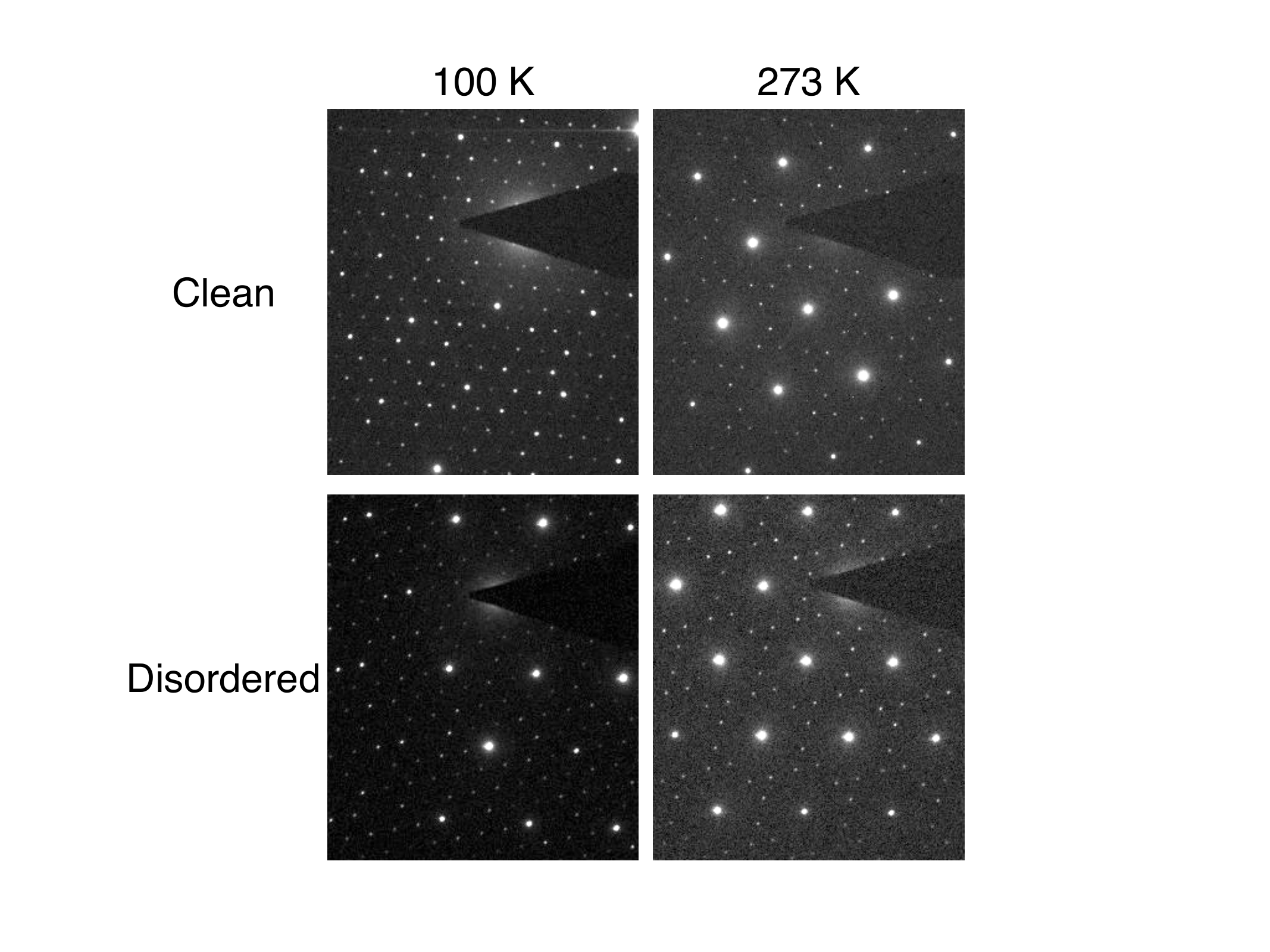}
\vspace{-1cm}
\caption{TEM diffraction images of a clean and disordered samples in the CCDW phase (left panels) and in the NCCDW phase (right panels)}
\label{fig:TEM}
\end{figure}

\noindent \emph{\underline{TEM:}}
Transmission Electron Microscopy (TEM) probes the periodic structure in the material and is ideal for detecting the presence of the electronic CDW modulation in the sample, and identifying the different phases of TaS$_2$. The CCDW-NCCDW transition can be seen in the clean sample in Fig.~\ref{fig:TEM} (top panels), above the transition the brightest spots correspond to the diffraction from the undistorted hexagonal Ta sheets, and the larger dim hexagons are associated with the formation of the "star-of-David" unit structure. Below the transition the stars interlock and tile the entire Ta sheet, yielding the new diffraction peaks below the transition.  We observed exactly the same behavior in the disordered samples, as shown Fig.~\ref{fig:TEM} (bottom panels). The addition of Cu did not suppress the CCDW phase.
  The electron diffraction images were acquired using a FEI Tecnai T20 Transmission Electron Microscope (TEM), using a  Gatan double-tilt cryo-stage (liquid N$_2$). The samples were cooled down to 100K and then warmed up back to room temperature. 
  The TEM images were taken with a spot size of about a micron and show sharp diffraction spots, this suggests that if intercalation produces domains, they must be at least as large as the spot size.
  
\section{Resistivity of pseudogapped metal}

Here we present a simple analysis for understanding the unusual low temperature behavior of resistivity $\rho(T)$ in the pseudogapped metallic phase. As seen in Fig. 2(g) of the main text, $\rho(T)$ has a negative slope as a function of temperature which is unexpected for typical metals.  As we discuss below, resistivity increasing with decreasing temperature and eventually saturating at very low temperatures is indeed a signature of pseudogapped metals. It is important to note hat the resistivity saturates at sufficiently low temperature for the pseudogapped metal while it diverges for a true insulator. 
Such a negative coefficient of resistivity has also been reported in the context of very high mobility (mean free path comparable to system size) suspended graphene samples with the chemical potential tuned close to the Dirac point \cite{kim_2008}. 

\begin{figure}[ht!]
\centering
\includegraphics[width=6cm]{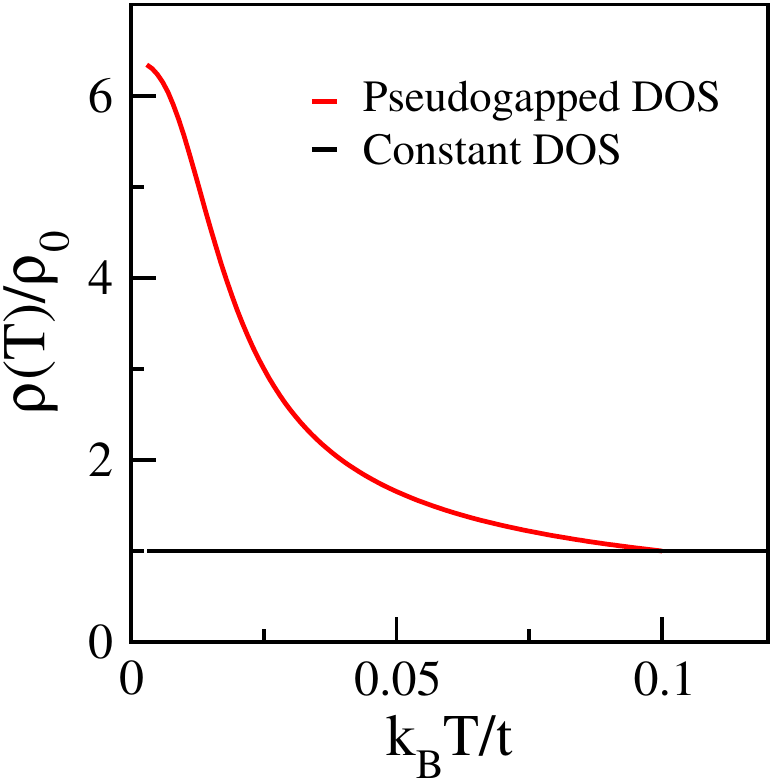}
\caption{Temperature dependence of resistivity for a pseudogapped metal (red) and normal metal (black).}
\label{fig:rho}
\end{figure}

The conductivity within the relaxation time approximation is given by \cite{AM}

\begin{equation}
  \sigma_{xx}  =  2e^2 \int \frac{d^3k}{(2\pi)^3} v_{k_x}^2  \tau_k \left( - \frac{\partial f_k}{\partial \epsilon_k}\right)
\end{equation}
Since the resistivity in these highly disordered materials is dominated by impurity scattering, $\tau_k$  is independent of temperature well below the Debye temperature $\Theta_D\approx 172K$ for 1T-TaS$_2$ \cite{debye_temp}.
The phonon contribution to resistivity scales as $(\textrm{T}/\Theta_D)^5$ which is negligible at low temperatures of T~5K and even at T=50K is of order $10^{-3}$.
Also at low temperatures only states close to the Fermi energy $\epsilon_F$ contribute to electrical transport, 
so we can replace $v_k$  by $v_F$  and $\tau_k$  by $\tau_F$. So, eqn (1) becomes
\begin{equation}
\sigma_{xx}  =  2e^2 v_{F}^2 \tau_{F} \int d\epsilon g(\epsilon) \left(-\frac{\partial f_\epsilon}{\partial \epsilon}\right)
\end{equation}
where $g(\epsilon)$ is the density of states (DOS). The behavior of the temperature-dependent conductivity $\sigma(T)$ is dictated by the energy dependence of the density of states $g(\epsilon)$, as shown in the Fig. \ref{fig:rho}. 

For a normal metal with a constant DOS, the resistivity $\rho(T)$ is independent of temperature. In sharp contrast, $\rho(T)$ decreases with increasing temperature for a pseudogapped metal as found in Cu intercalated 1T-TaS2 or in graphene with  strongly energy-dependent $g(\epsilon)$. In Fig. \ref{fig:rho} we have used the same DOS as in Fig. 4(a) of main text (with V=6t) for the pseudogapped metal. 

The negative slope of resistivity is expected to persist as long as the temperature is less than the pseudogap scale which for Cu-interacalated 1T-TaS$_2$ is about 0.1eV or 1100K (see Fig 3(c) of main text). Of course in the real material, phonons become relevant at higher temperatures and these additional scattering mechanisms must be included to describe the resistivity at higher temperatures.
In addition, the approximations to calculate the resistivity are only valid below the melting temperature T$\approx$180K of the charge density wave state.

As mentioned above, the main difference between a pseudogapped metal and a true insulator is that the resistivity of a pseudogapped metal saturates at sufficiently low temperature while that of an insulator diverges at T=0. Our estimate for the saturation temperature is about 0.5K which is below the low temperature limit of the experiments performed so far but should be accessible in future experiments.

\bibliography{references}
\bibliographystyle{apsrev}